\documentclass[a4paper]{article}
\usepackage[margin=25mm]{geometry}
\usepackage{amsmath,amssymb}
\usepackage{graphicx}
\usepackage{algorithm}
\usepackage{algorithmic}
\usepackage[normalsize]{subfigure}
\usepackage{setspace}
\usepackage{wrapfig}

\title{
\vspace{-2em}
Unbounded Slice Sampling
}

\author{
Daichi Mochihashi \\
The Institute of Statistical Mathematics \\
{\tt daichi@ism.ac.jp}
}

\date{
2020-1-5 (Sun)
\vspace{-1em}
}

\def\R{\mathbb{R}}
\def\N{\mathcal{N}}
\def\unit{\ensuremath{[\kern1.1pt 0,1)}}

\begin{document}

\maketitle

\begin{abstract}
Slice sampling is an efficient Markov Chain Monte Carlo algorithm to sample
from an unnormalized density with acceptance ratio always $1$.
However, when the variable to sample is unbounded, its ``stepping-out''
heuristic works only locally, making it difficult to uniformly explore
possible candidates. 
This paper proposes a simple change-of-variable method to slice sample
an unbounded variable equivalently from $\unit$.
\end{abstract}

\section{Introduction}

Slice sampling~\cite{neal03slice} is one of the Markov Chain Monte Carlo
(MCMC) methods to sample from one-dimensional distribution.
Because its acceptance probability is always $1$ as opposed to famous
Metropolis-Hastings algorithm~\cite{gilks96mcmc}, 
slice sampling is widely employed in modern statistics and machine learning 
problems, especially for sampling 
hyperparameters of a statistical model where each hyperparameter is often
univariate but has a nontrivial likelihood surface.

When the variable to sample is not bounded, an algorithm called
``stepping-out''~\cite{mackay03itila} is usually employed to adaptively adjust
the interval to sample from based on the current value of the variable.
However, this ``stepping-out'' works only locally, thus will be potentially 
trapped to local modes in multimodal likelihoods. In contrast, in this paper 
we propose a simple method to uniformly sample from an unbounded variable
through an appropriate reparametrization with a unit interval $\unit$.

\section{Slice sampling}

\begin{figure}[tbp]
 \centering
 \subfigure[Sigmoid map $\displaystyle p = \frac{1}{1+e^{-x}}$]{
  \includegraphics[scale=0.42]{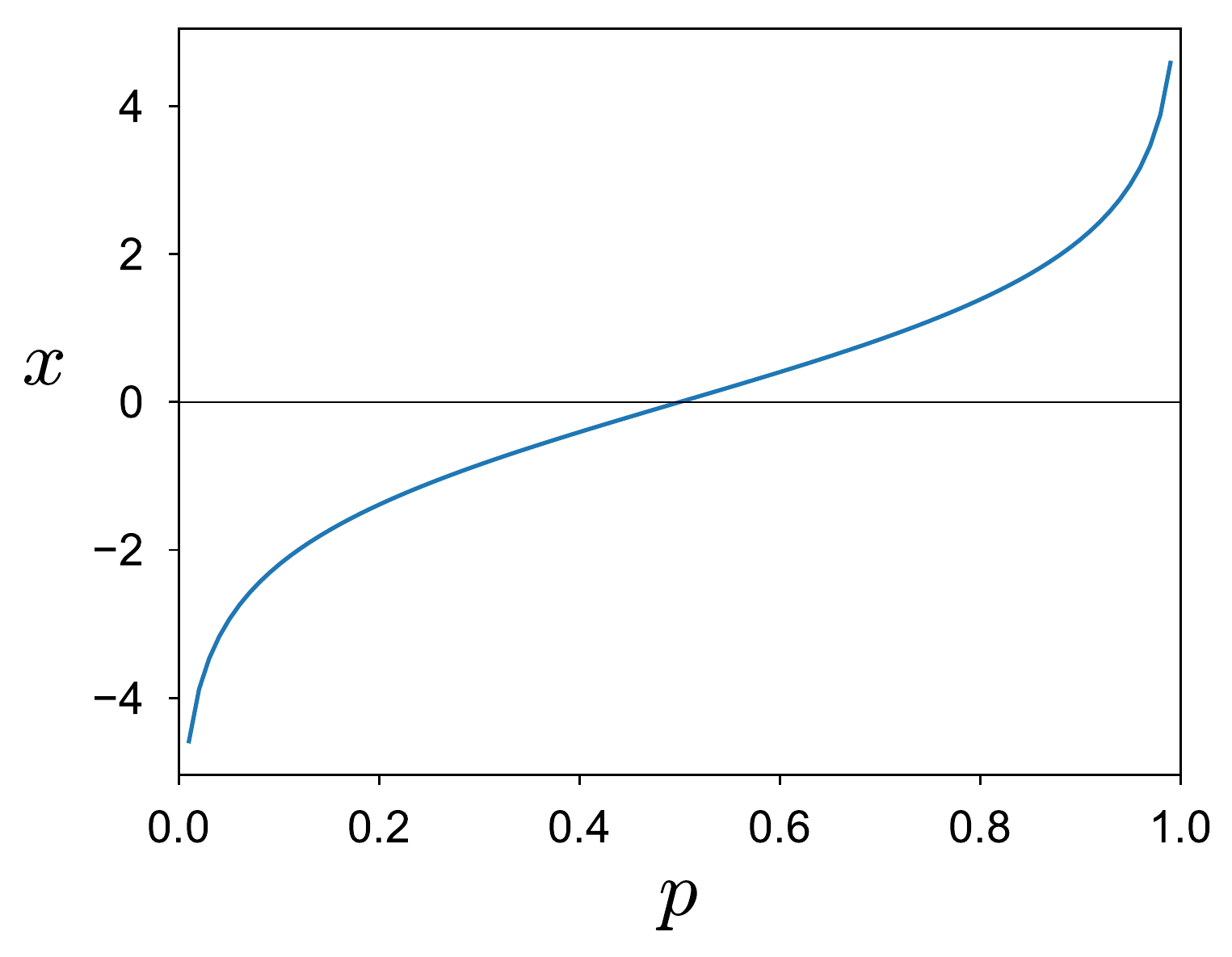}
  \label{figure:sigmoid-map}
 }
 \subfigure[Logit map $\displaystyle x = \frac{p}{1-p}$]{
  \includegraphics[scale=0.42]{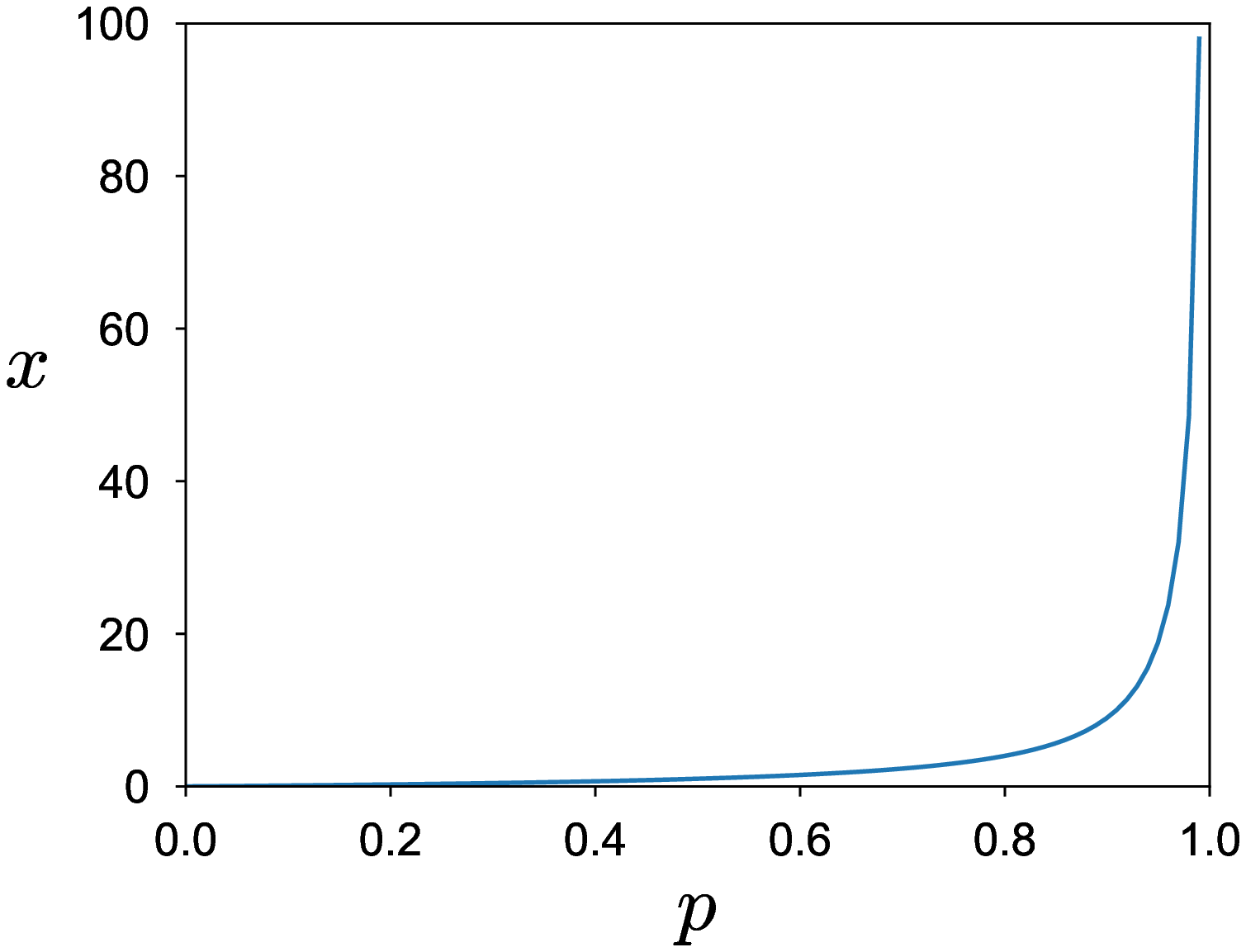}
  \label{figure:logit-map}
 }
 \caption{One-to-one map between $p \in \unit$ and unbounded $x$.}
\end{figure}

Slice sampling iteratively draws $x$ from a probability distribution
\begin{equation}
 p(x) = \frac{f(x)}{Z}
\end{equation}
without knowing the normalizing constant 
$\displaystyle Z \!=\! \int\! f(x)dx$ by the following general
algorithm, starting from an initial value $x^{(0)}$:
\begin{center}
\begin{minipage}{0.9\textwidth}
\begin{algorithm}[H]
\caption{General slice sampling}
\label{algo:general-slice-sampling}
\begin{algorithmic}[1]
 \FOR{$t = 1 \cdots T$\rule{0pt}{1em}}
 \STATE{$\ell = f(x^{(t-1)})$}
 \STATE{$\rho \sim {\rm Uniform}[0,\ell)$}
 \STATE{$x^{(t)} \sim {\rm Uniform}(\{ x: f(x) > \rho \})$}
 \ENDFOR
\end{algorithmic}
\end{algorithm}
\end{minipage}
\vspace{0.5em}
\end{center}

In practice, Step $4$ of the Algorithm~\ref{algo:general-slice-sampling} 
is not trivial, and usually
``stepping-out'' heuristic~\cite{neal03slice} is employed to determine the 
interval from which $x$ is uniformly drawn,
starting from an initial interval enclosing $x$ 
that is adaptively expanded.

However, the efficiency of ``stepping-out'' depends on setting the initial
interval whose appropriate scale for the problem is not known in advance.
Moreover, it works only locally: if $f(x)$ has multiple and distant modes,
it tends to wander around a single mode and will spend a very long time to
explore different modes that might have higher probabilities.

On the other hand, if $x$ is bounded and known to always reside in the
interval $[{\it st}, {\it ed}]$, slice sampling is very easy 
using the following algorithm:
\begin{center}
\vspace{-0.5em}
\begin{minipage}{0.9\textwidth}
\begin{algorithm}[H]
\caption{~Bounded slice sampling}
\label{algo:bounded-slice-sampling}
\begin{spacing}{0.95}
\begin{algorithmic}[1]
 \FOR{$t = 1 \cdots T$\rule{0pt}{1em}}
 \STATE{$\ell = f(x^{(t-1)})$}
 \STATE{$\rho \sim {\rm Uniform}[0,\ell)$}
 \WHILE{true}
  \STATE{$x \sim {\rm Uniform}[{\it st},{\it ed}]$
  \qquad /* generate a candidate */
  }
  \renewcommand{\algorithmicelse}{\textbf{else}{%
  \hspace{9em}/* modify interval */}}
  \IF{$f(x) > \rho$}
   \STATE{$x^{(t)} = x$; break}
  \ELSE
  \renewcommand{\algorithmicelse}{\textbf{else}}
   \IF{$x < x^{(t-1)}$}
    \STATE{${\it st} := x$}
   \ELSE
    \STATE{${\it ed} := x$}
   \ENDIF
  \ENDIF
 \ENDWHILE
 \ENDFOR
\end{algorithmic}
\end{spacing}
\end{algorithm}
\end{minipage}
\end{center}

This is an efficient binary search that quickly draws $x$ uniformly
over the region $[{\it st},{\it ed}]$ where $f(x) > \rho$.

\section{Unbounded slice sampling}

Algorithm~\ref{algo:bounded-slice-sampling} is effective for a bounded $x$ 
that is known to reside in $[{\it st},{\it ed}]$.
However, even when $x$ is unbounded, we can sample $x$
through an one-to-one map between $\unit$ and $\R$.

\paragraph{General unbounded variate:}

For example, using a sigmoid map
\begin{align}
 p &= \frac{1}{1+e^{-x}} 
 \label{eqn:p-from-x}
 \\
 \text{\it i.e.}\qquad
 x &= - \log \left( \frac{1}{\,p\,} - 1 \right)
 \label{eqn:x-from-p}
\end{align}
for $p \in \unit$ shown in Figure~\ref{figure:sigmoid-map}, 
for each $x \in \R$ we can associate $p \in \unit$ and vice versa.

\medskip

Since
\begin{equation}
 \frac{dx}{dp} = \frac{1}{p(1-p)}~,
\end{equation}
we can alternatively sample $p$ in place of $x$ using 
the exchange of variables:
\begin{equation}
 f(x)dx = f(x)\frac{dx}{dp}dp 
 = f\left(- \log \left( \frac{1}{\,p\,} - 1 \right)\right)
   \frac{1}{p(1-p)} dp
 \label{eqn:exchanged-density}
\end{equation}

Using \eqref{eqn:exchanged-density} for the 
Algorithm~\ref{algo:bounded-slice-sampling} in place of $f(x)$ can sample
$p \in \unit$, from which we can recover $x$ by the relationship
\eqref{eqn:x-from-p}.

\paragraph{Positive variate:}

When $x > 0$, for example we can exploit a map
\begin{equation}
 x = \frac{p}{1-p}
 \label{eqn:positive-x-from-p}
\end{equation}
therefore
\begin{equation}
 \frac{dx}{dp} = \frac{1}{(1-p)^2}~,
\end{equation}
leading
\begin{equation}
 f(x)dx = f(x)\frac{dx}{dp}dp = f\left( \frac{p}{1-p} \right)
          \frac{1}{(1-p)^2} dp
\end{equation}
to sample $p \in \unit$ and obtain $x$ by the 
relationship~\eqref{eqn:positive-x-from-p}.

\medskip

Using this reparameterization, we can slice sample an unbounded $x$
essentially
by a ``probabilistic binary search'' always on $\unit$.
We included sample of MATLAB and C codes for our unbounded slice sampling 
in Appendix B and C.

\medskip

\begin{wrapfigure}[15]{r}{0.4\textwidth}
 \centering
 \vspace{-3.2em}
 \includegraphics[scale=0.5]{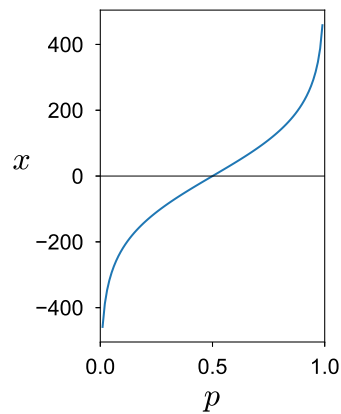}
 \vspace{-1em}
 \caption{Modified sigmoid map $\displaystyle p = \frac{1}{1+e^{-x/100}}$.}
 \label{figure:sigmoid-A}
\end{wrapfigure}

In practice, sometimes we cannot compute a corresponding $p$ for very large
(or small) $x$ using equation~\eqref{eqn:p-from-x}.
Therefore, instead of \eqref{eqn:p-from-x} we can use
\begin{equation}
 p = \sigma(x/A) = \frac{1}{1+e^{-x/A}}
 \label{eqn:sigmoid-map-modified}
\end{equation}
for some constant $A > 0$ as shown in Figure~\ref{figure:sigmoid-A}.
While every $x \in \R$ is mathematically equivalent, notice that usually
the parameter to sample from a statistical model roughly\footnote{
This is important, because we do not enforce a strict interval to sample $x$
in this paper.
}
resides in this regime; even if $x$ will exceed over this regime,
we can safely rescale $x$ to fit into a decent interval.
We empirically confirmed that $A=100$ usually works fine for 
$-1000 < x < 1000$
(see Section~\ref{sec:experiments}).

\section{Experiments}
\label{sec:experiments}

We conducted some experiments to confirm that our unbounded slice sampler
correctly samples from a given distribution.
Using the C code shown in Appendix~\ref{figure:c-slice-sampling},
we can easily sample from $p(x)$.

\begin{figure}[tbp]
 \centering
 \subfigure[$f(x) = \exp(-x (x\!-\!1) (x\!-\!2) (x\!-\!3.5))$]{
  \includegraphics[scale=0.42]{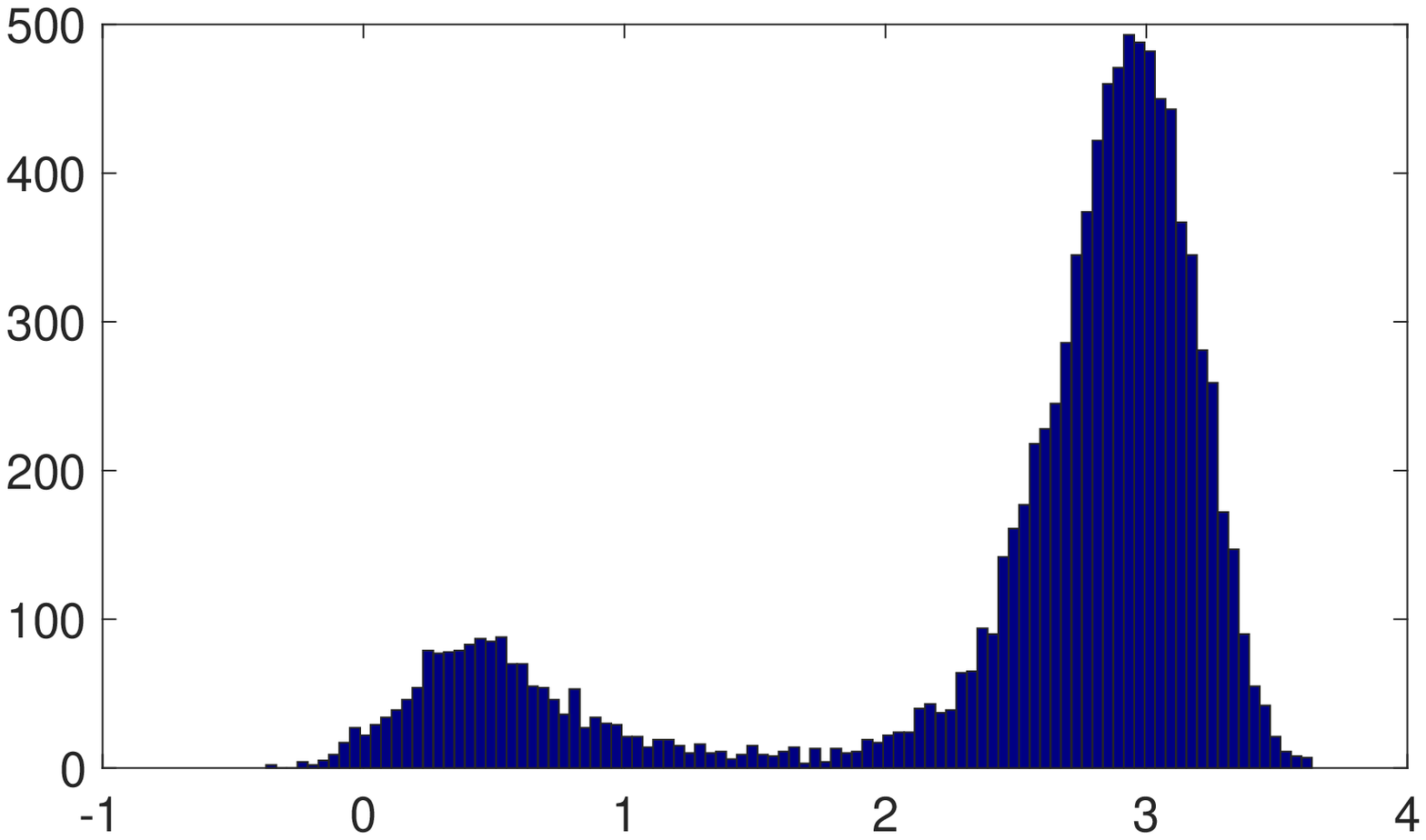}
 }~
 \subfigure[$f(x) = \exp(- (x \!-\! 500)^2 / 10)$]{
  \includegraphics[scale=0.42]{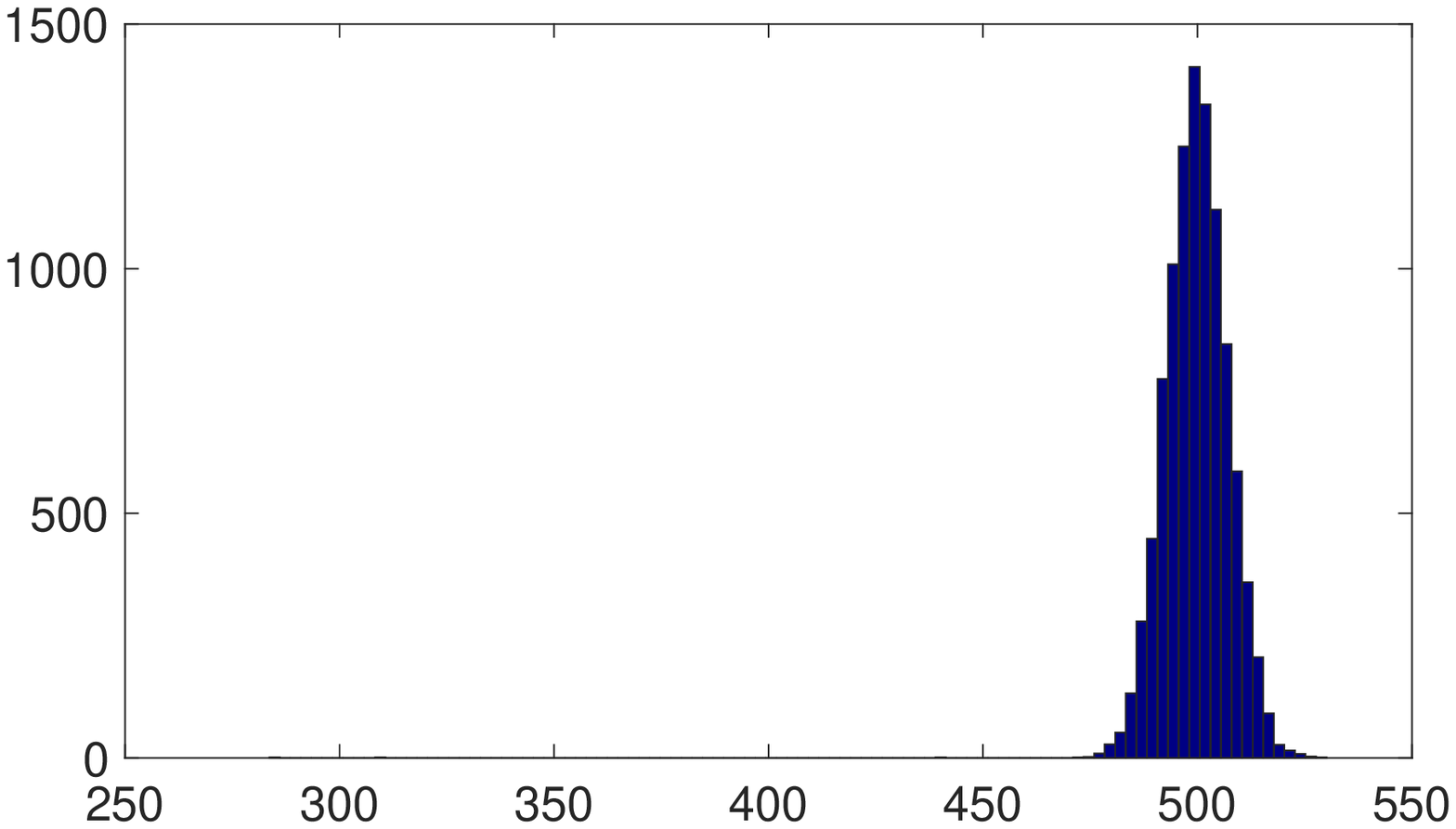}
 }\\[1em]
 \subfigure[$f(x) = \exp(- (x \!-\! 1000)^2 / 100)$]{
  \includegraphics[scale=0.42]{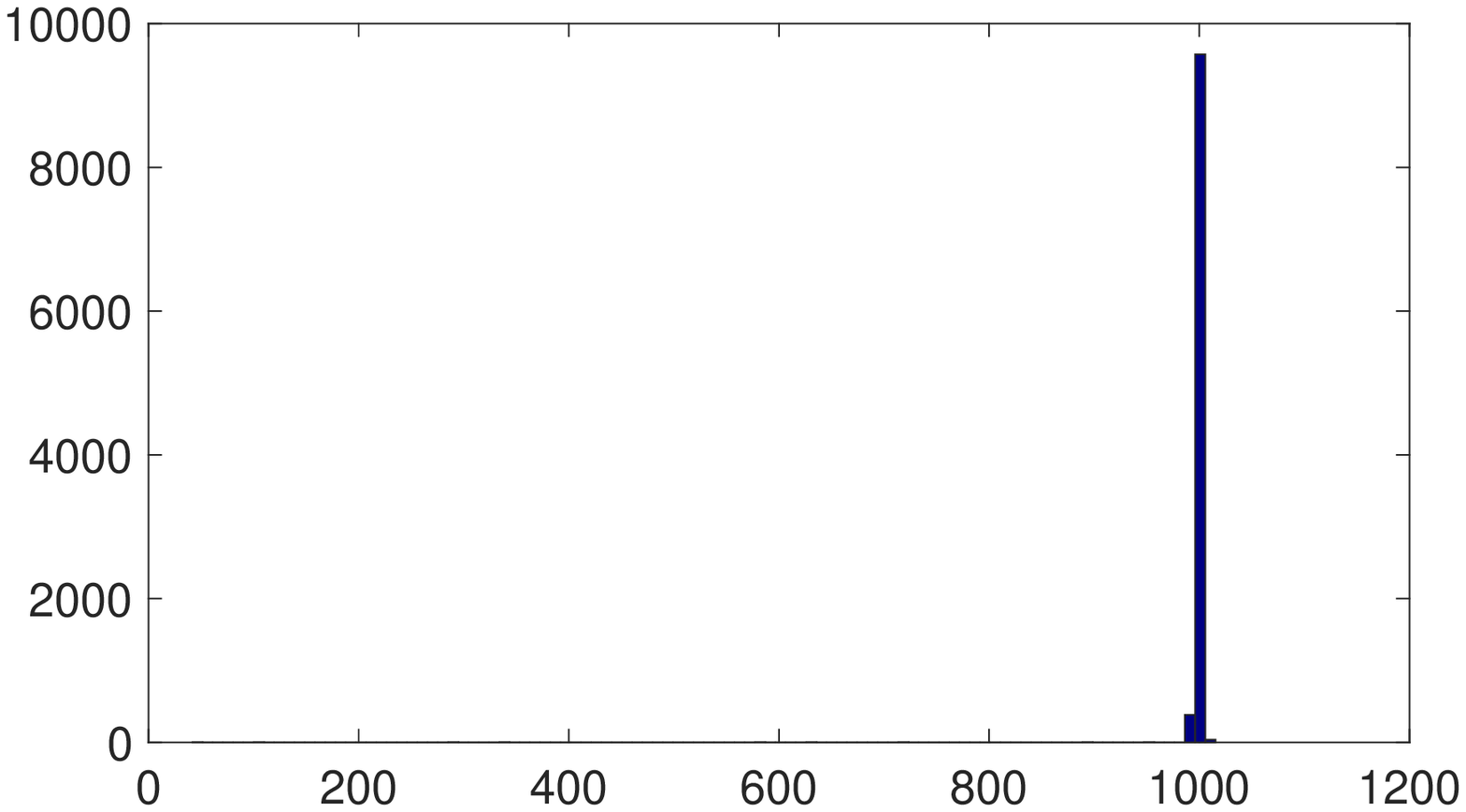}
  \label{fig:distant-mode}
 }
 \caption{Unbounded slice sampling of $x \in \R$.}
 \label{figure:slice-sample-test}
\end{figure}

For a general unbounded case $x \in \R$, Figure~\ref{figure:slice-sample-test}
shows the plots of sampled $x$ from
(a) $f(x) = \exp(- (x \!-\! 500)^2 / 10)$,
(b) $f(x) = \exp(- (x \!-\! 500)^2 / 10)$, and
(c) $f(x) = \exp(- (x \!-\! 1000)^2 / 100)$.
Note that $p(x)$ will be high on different regimes for each case:
roughly $0<x<4$ for (a), $x \approx 500$ for (b), and $x \approx 1000$ 
for (c), while these regimes are not known in advance, and they have different
variances.
Figure~\ref{figure:slice-sample-test} clearly shows we can correctly sample
from $p(x)$, even when $x$ is very large using the map
\eqref{eqn:sigmoid-map-modified} with $A\!=\!100$.
The average step of ``binary search'', i.e. the number of function evaluations
for the loop in Step 4 of the Algorithm~\ref{algo:bounded-slice-sampling},
is $11.44$ for (a), $16.48$ for (b),
and $9.34$ for (c).


\begin{figure}[htb]
 \centering
 \subfigure[$f(x) = x^4 e^{-x}$~$(={\rm Gamma}(5,1))$]{
  \includegraphics[scale=0.42]{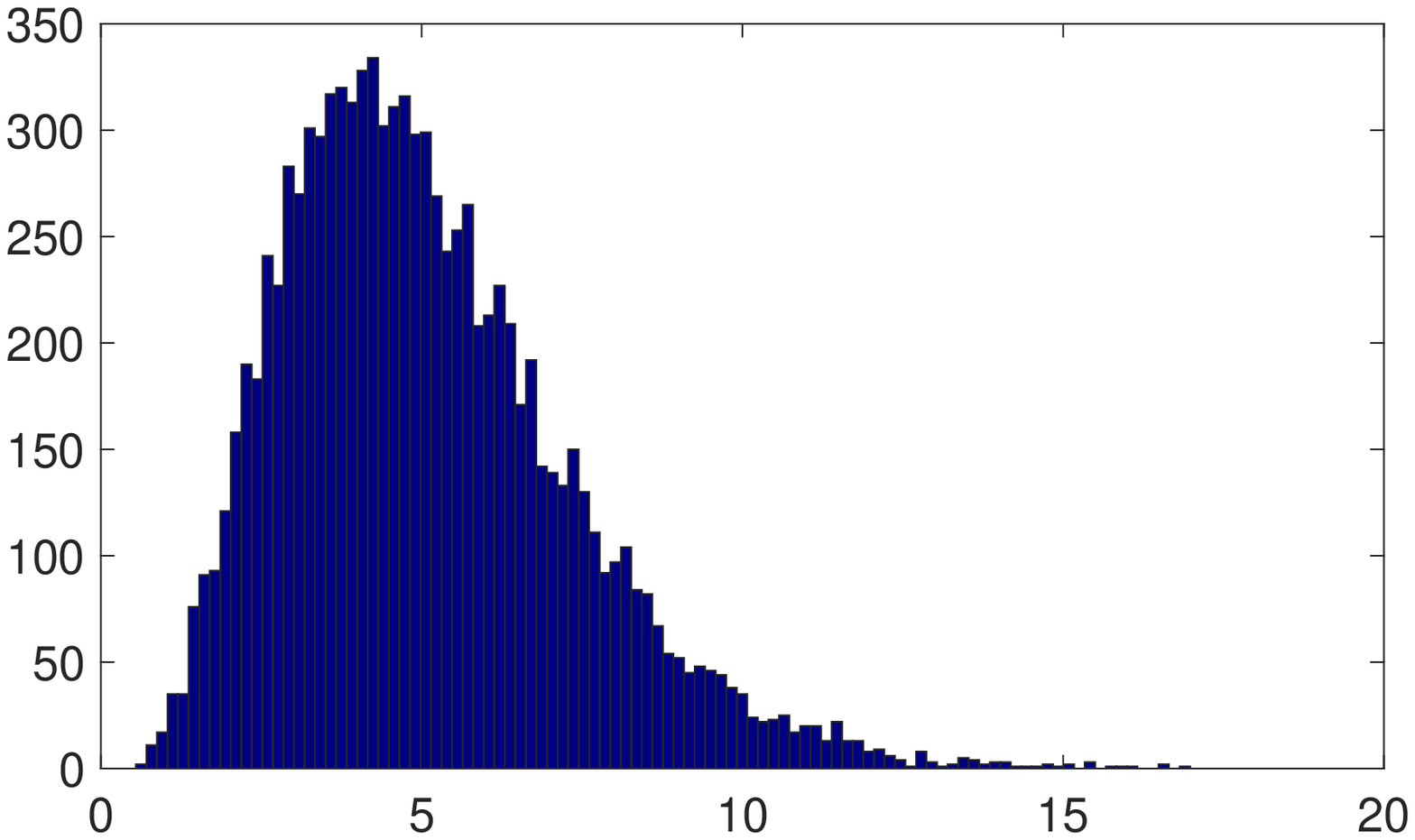}
 }~
 \subfigure[$f(x) = \exp(-(x\!-\!1000)^2/100)$]{
  \includegraphics[scale=0.42]{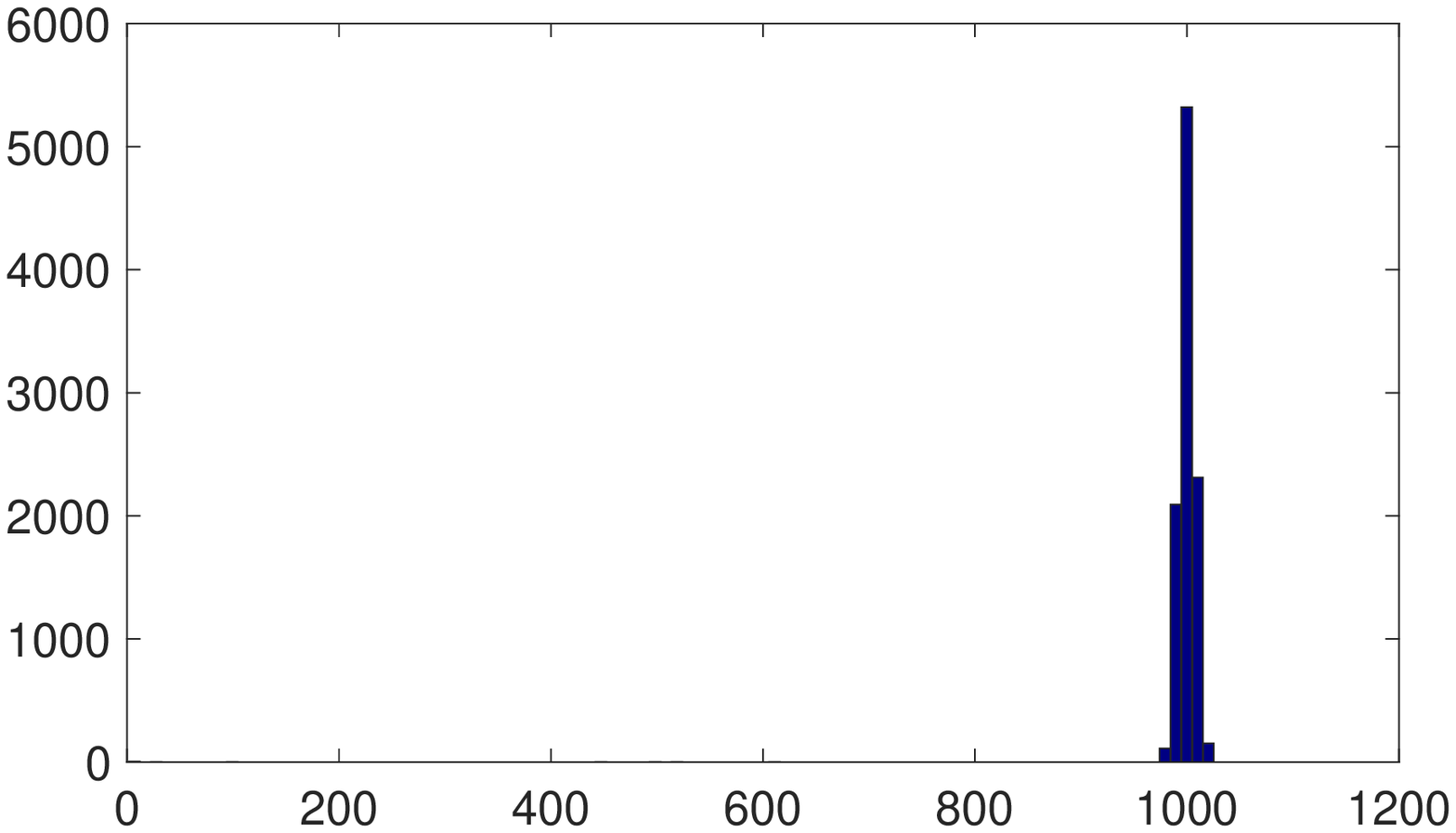}
 } 
 \caption{Unbounded slice sampling of $x > 0$.}
 \label{figure:slice-test-positive}
\end{figure}

For a positive case $x \in \R^+$, 
Figure~\ref{figure:slice-test-positive} shows the results for
(a) $f(x) = x^4 e^{-x}$ and (b) $f(x) = \exp(-(x\!-\!1000)^2/100)$;
(a) means $p(x) \sim {\rm Gamma}(5,1)$.

\paragraph{Comparison with stepping-out}

\begin{figure}[t]
 \centering
 \subfigure[Unbounded slice sampling]{
  \includegraphics[scale=0.41]{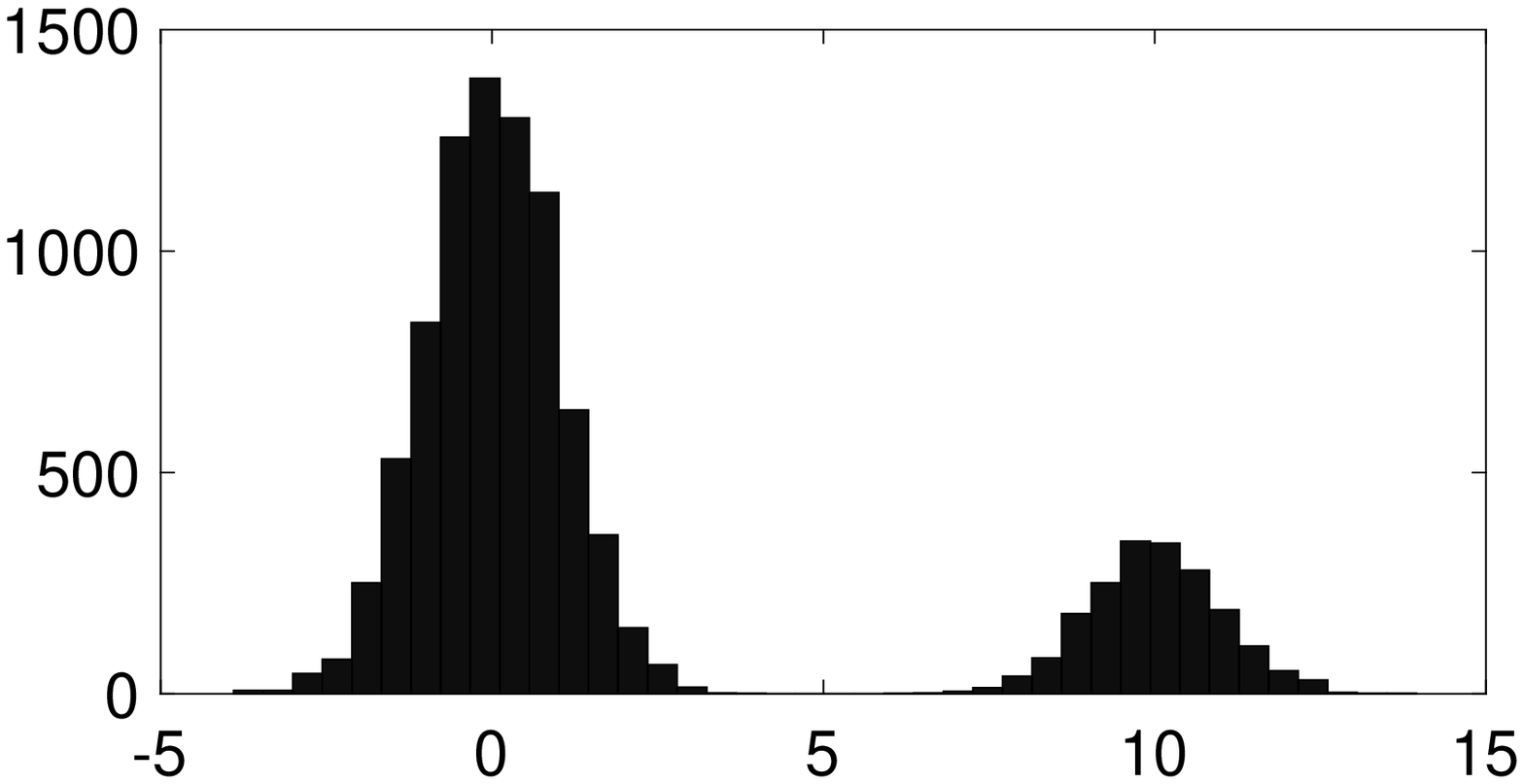}
 }
 \subfigure[Stepping-out algorithm (width=$1$)]{
  \includegraphics[scale=0.41]{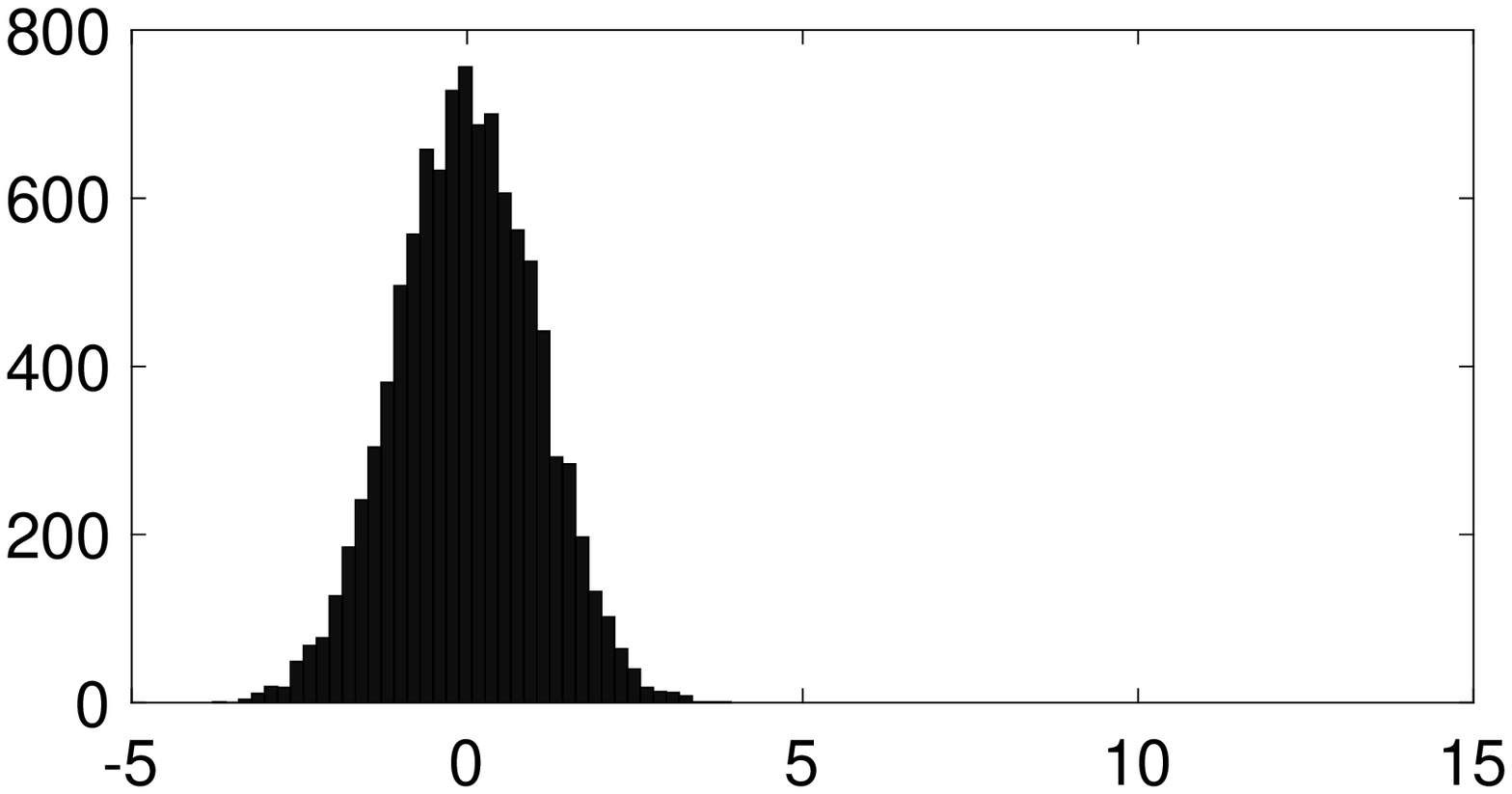}
 }
 \caption{Comparison with ``stepping-out'' on a Gaussian mixture model
 $p(x) = 0.8 \N(x|0,1^2) + 0.2\N(x|10,1^2)$ using $10,000$ samples.}
 \label{figure:comparison}
\end{figure}

To confirm the advantage of our unbounded slice sampling,
we also conducted an experiment to sample from a multimodal distribution.
Specifically, Figure~\ref{figure:comparison} shows a draw of $10,000$ samples
from a Gaussian mixture model
$p(x) = 0.8 \N(x|0,1^2) + 0.2\N(x|10,1^2)$
using unbounded slice sampler and ``stepping-out'' algorithm, respectively.

Clearly, ``stepping-out'' is stuck into the first mode because the sampling
starts from $x=1$, and this tendency remained the same over multiple 
simulations.
Moreover, because ``stepping-out'' linearly increases the interval to
sample from, sampling from a distant density shown in 
Figure~\ref{fig:distant-mode} required about $2,000$ function evaluations
at first to reach the mode, while unbounded slice sampling required only
$9.34$ as described before, thanks to the efficient ``probabilistic binary
search'' on $\R$.

\bibliographystyle{unsrt}
\bibliography{unbounded}

\begin{thebibliography}{1}

\bibitem{neal03slice}
Radford~M. Neal.
\newblock {Slice sampling}.
\newblock {\em {Annals of Statistics}}, pages {705--741}, {2003}.

\bibitem{gilks96mcmc}
W.~R. Gilks, S.~Richardson, and D.~J. Spiegelhalter.
\newblock {\em {Markov Chain Monte Carlo in Practice}}.
\newblock {Chapman \& Hall / CRC}, {1996}.

\bibitem{mackay03itila}
David J.~C. MacKay.
\newblock {\em {Information Theory, Inference, and Learning Algorithms}}.
\newblock {Cambridge University Press}, {2003}.

\end{thebibliography}

\newpage

\appendix
\addcontentsline{toc}{section}{Appendices}
\section*{Appendix}

\section{C code of easy slice sampling}

\begin{spacing}{0.85}
\begin{verbatim}
#include "slice.h"

double
f (double x, void *arg)
{
        return - x * (x - 1) * (x - 2) * (x - 3.5);
}

main () {
        int i, N = atoi (argv[1]);
        double x = 1;

        for (i = 0; i < N; i++) {
                x = slice_sample (x, f, NULL);
                printf("%lf\n", x);
        }
}
\end{verbatim}
\end{spacing}
\label{figure:c-slice-sampling}

\section{C code for unbounded slice sampling}

{\small
\begin{spacing}{0.85}
\begin{verbatim}
/*
 *    slice.c
 *    Unbounded slice sampling with ease.
 *    $Id: unbounded.tex,v 1.17 2020/01/03 01:23:08 daichi Exp $
 *
 */
#include <stdio.h>
#include <stdlib.h>
#include <math.h>
#include "random.h"

static double A = 100.0;
typedef double (*likfun)(double x, void *args);

static double
expand (double p)        /* p -> x */
{
        return - A * log (1 / p - 1);
}

static double
shrink (double x)        /* x -> p */
{
        return 1 / (1 + exp(- x / A));
}

static double
expandp (double p)        /* p -> (x > 0) */
{
        return p / (1 - p);
}

static double
shrinkp (double x)        /* (x > 0) -> p */
{
        return  x / (1 + x);
}

double
slice_sample (double x, likfun loglik, void *arg)
{
        double st = 0, ed = 1;
        double r, rnew, slice, newlik;
        int iter, maxiter = 1000;

        r = shrink (x);
        slice = (*loglik)(x, arg) - log (A * r * (1 - r)) + log (RANDOM);
        
        for (iter = 0; iter < maxiter; iter++)
        {
                rnew = unif (st, ed);
                newlik = (*loglik)(expand(rnew), arg)
                                        - log (A * rnew * (1 - rnew));

                if (newlik > slice)
                        return expand (rnew);
                else if (rnew > r)
                        ed = rnew;
                else if (rnew < r)
                        st = rnew;
                else
                        return x;
        }
        fprintf(stderr, "slice_sample: max iteration %d reached.\n", maxiter);
        return x;
}

double
slice_sample_positive (double x, likfun loglik, void *arg)
{
        double st = 0, ed = 1;
        double r, rnew, slice, newlik;
        int iter, maxiter = 1000;

        r = shrinkp (x);
        slice = (*loglik)(x, arg) - 2 * log (1 - r) + log (RANDOM);
        
        for (iter = 0; iter < maxiter; iter++)
        {
                rnew = unif (st, ed);
                newlik = (*loglik)(expandp(rnew), arg) - 2 * log (1 - rnew);

                if (newlik > slice)
                        return expandp (rnew);
                else if (rnew > r)
                        ed = rnew;
                else if (rnew < r)
                        st = rnew;
                else
                        return x;
        }
        fprintf(stderr, "slice_sample: max iteration %d reached.\n", maxiter);
        return x;
}

#if 1

double
f (double x, void *arg)
{
        return - x * (x - 1) * (x - 2) * (x - 3.5);
}

double
g (double x, void *arg)
{
        return - 10 * (x + 1000) * (x + 1000);
}

double
h (double p, void *arg)
{
        double a = 2, b = 3;
        return (a - 1) * log (p) + (b - 1) * log (1 - p);
}

int
main (int argc, char *argv[])
{
        int i, N = atoi(argv[1]);
        double x = 0.5;

        for (i = 0; i < N; i++)
        {
                x = slice_sample (x, g, NULL);
                // x = slice_sample_positive (x, f, NULL);
                printf("%.4lf\n", x);
        }
}

#endif
\end{verbatim}
\end{spacing}
}
\vspace{-1em}

\section{MATLAB code for unbounded slice sampling}

{\small
\begin{spacing}{0.9}
\begin{verbatim}
function xnew = slice_sample (x,likfun,varargin)
% xnew = slice_sample (x,likfun,varargin)
% Unbounded slice sampling in MATLAB.
% $Id: unbounded.tex,v 1.17 2020/01/03 01:23:08 daichi Exp $
global A;
A  = 100;
st = 0; ed = 1;
maxiter = 1000;
% function body
r = shrink (x);
slice = feval (likfun,x,varargin{:}) - log (A * r * (r - 1)) + log (rand());
% slice sampling
for iter = 1:maxiter
  rnew = unif (st,ed);
  xnew = expand (rnew);
  newlik = feval (likfun,xnew,varargin{:}) - log (A * rnew * (rnew - 1));
  if (newlik > slice)
    return;
  elseif (rnew > r)
    ed = rnew;
  elseif (rnew < r)
    st = rnew;
  else
    return;
  end
end
fprintf(stderr,'slice_sample: max iteration %d reached\n',maxiter);
return;

function p = shrink(x)	% x -> p
global A;
p = 1 / (1 + exp (- x / A));

function x = expand(p)	% p -> x
global A;
x = - A * log (1 / p - 1);

function xnew = slice_sample_positive (x,likfun,varargin)
% xnew = slice_sample_positive (x,likfun,varargin)
% Unbounded slice sampling in MATLAB. (only positive)
% $Id: unbounded.tex,v 1.17 2020/01/03 01:23:08 daichi Exp $
st = 0; ed = 1;
maxiter = 1000;
% function body
r = shrinkp (x);
slice = feval (likfun,x,varargin{:}) - 2 * log (1 - r) + log (rand());
% slice sampling
for iter = 1:maxiter
  rnew = unif (st,ed);
  xnew = expandp (rnew);
  newlik = feval (likfun,xnew,varargin{:}) - 2 * log (1 - rnew);
  if (newlik > slice)
    return;
  elseif (rnew > r)
    ed = rnew;
  elseif (rnew < r)
    st = rnew;
  else
    return;
  end
end
fprintf(stderr,'slice_sample: max iteration %d reached\n',maxiter);
return;

function p = shrinkp (x)	% (x > 0) -> p
p = x / (1 + x);

function x = expandp (p)	% p -> (x > 0)
x = p / (1 - p);

\end{verbatim}
\end{spacing}
}

\end{document}